\documentstyle[11pt,newpasp,twoside]{article}
\def\edcomment#1{\iffalse\marginpar{\raggedright\sl#1\/}\else\relax\fi}
\reversemarginpar

\begin{document}
\title{Stream-fed Accretion in Intermediate Polars}

\author{Coel Hellier} 

\affil{Astrophysics Group, Keele University, Staffs ST5 5BG, U.K.}

\begin{abstract}
I review the observational evidence for stream-fed accretion in
intermediate polars. Recent work on the discless system V2400~Oph
confirms the pole-flipping model of stream-fed accretion, but this
applies only to a minority of the flow. The bulk of the flow is in the
form of blobs circling the white dwarf, a state which might have been a
precursor to disc formation in other IPs. I also discuss
work on the systems with anomalously long spin periods, V1025~Cen and
EX~Hya. There are arguments both for and against stream-fed accretion
in V1025~Cen, and further work is necessary before reaching a conclusion
about this system. 
\end{abstract}

\section{Introduction}
Early work on intermediate polars (IPs), taking a lead from studies of
DQ~Her, tended to assume the presence of an accretion disc. The
accretion stream would feed the outer disc, as in non-magnetic
systems, and the magnetic field would disrupt the inner disc and so
channel the accretion along field lines onto magnetic polecaps.
Hameury, King, \&\ Lasota (1986) and Lasota \&\ King (1991) presented
theoretical arguments against this idea. They suggested that most IPs
were discless, with the accretion stream falling until it encountered
the magnetosphere directly.  They argued that, observationally,
discless systems would be distinguished by X-ray modulations over the
orbital cycle, since in a stream-fed system the accretion sites would
lie `beneath' the stream, and so be localised in orbital phase.

Hellier (1991) and Hellier, Garlick, \&\ Mason (1993) pointed out that
X-ray orbital modulations can also result from obscuration of the
white dwarf by structure in an accretion disc, as seen in `dipping'
LMXBs, and so are not exclusive to discless systems.  We suggested
that a better diagnostic was the presence of an X-ray pulsation at the
beat frequency ($\omega$\,--\,$\Omega$) between the spin ($\omega$) and 
orbital ($\Omega$) frequencies.  This arises because the geometry changes
on the beat frequency when a magnetosphere (spinning at $\omega$)
rotates beneath a stream (orbiting at $\Omega$). Indeed we would
expect the stream to follow the field line involving
least deviation from the orbital plane, and so flow to the upper 
magnetic pole for half the beat cycle and then flip to the lower pole
for the remainder of the cycle.

Wynn \&\ King (1992) modelled the situation and confirmed that X-ray
modulations at $\Omega$ and $\omega$\,--\,$\Omega$ are characteristic of 
stream-fed accretion, but also found that power could be moved from 
$\omega$\,--\,$\Omega$ to 2$\omega$\,--\,$\Omega$ for certain 
combinations of the system inclination and the angle between the 
magnetic and spin axes.

Hellier (1991; 1992) analysed X-ray lightcurves to show that all
currently known IPs were dominated by X-ray pulsations at the spin
period, rather than the beat period, and so argued that they were all
disc-fed accretors. However, Buckley et\,al.\ (1995; 1997) then
discovered the {\it Rosat\/} source V2400~Oph, which
showed a strong X-ray pulsation at the 1003-sec beat period but none at
the 927-sec spin period, and thus was the first secure case of discless
accretion amongst the known IPs.

The 927-sec period was detected only in polarised light, and this raised
the following problem for deductions based on X-ray periodicities:
since many IPs show no polarised light, might we fail to detect the
spin period in a non-polarised discless system, and if so might we be
misinterpreting the beat cycles as spin cycles? Exactly this issue was
debated in the case of BG~CMi (Norton et\,al.\ 1992b; de Martino et\,al.\ 
1995; Hellier 1997).  New
spectroscopy of V2400~Oph, reported in Hellier \&\ Beardmore (2002),
is reassurring on this point.  Both the spin and the beat cycles are
seen easily in the emission lines (Fig.~1), and this would leave
little doubt as to their respective identities, even if we had no
polarimetry.  Thus V2400~Oph greatly strengthens the overall argument,
since the discovery of a system showing so clearly the predicted
hallmark of discless accretion confirms the validity of
distinguishing between disc-fed and stream-fed accretors using the 
spin and beat pulses in the X-ray lightcurve.

\begin{figure}[t]   % Fig 1
\vspace*{8cm}
\includegraphics{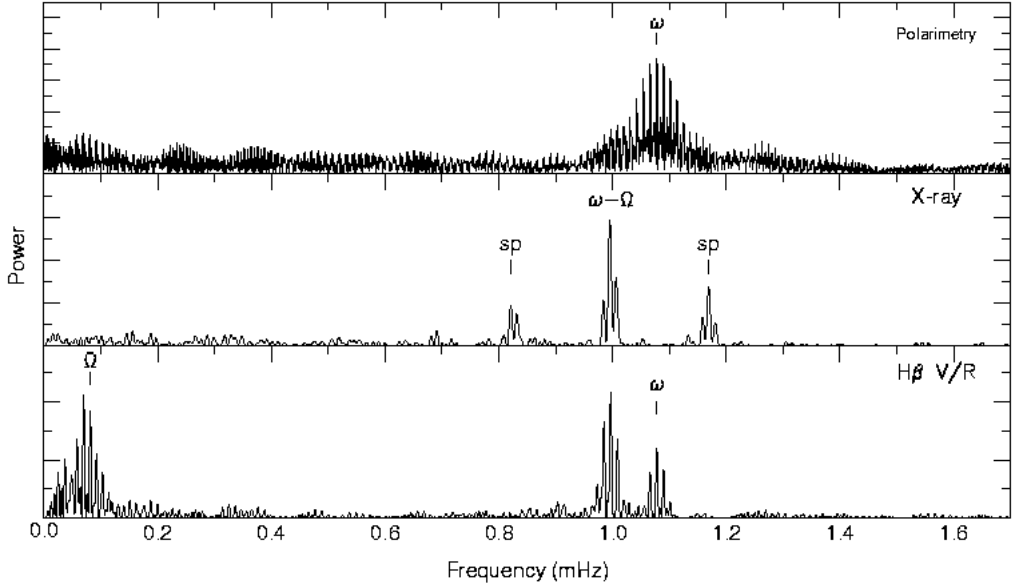}
\caption{Fourier transforms from V2400~Oph, comparing the polarised
light (data by David Buckley) with X-ray data and 
H$\beta$ V/R ratios. The spin, orbital and beat frequencies are marked
with the usual notation of $\omega$, $\Omega$ and $\omega$\,--\,$\Omega$,
respectively. Peaks marked `sp' are windowing caused by the spacecraft
orbit.}
\end{figure}

Since nearly all well-studied IPs show both spin and beat cycles in
the emission lines, we can have confidence in assigning cycles correctly
and thus in applying the argument to the whole class.

\section{Stream/disc overflow}
Even in systems where a spin-cycle pulsation dominates the X-ray
lightcurves, a weaker beat-cycle pulsation is often seen (e.g.\ Buckley
\&\ Tuohy 1989; Hellier 1991; Norton et\,al.\ 1992a; Hellier 1998). These 
have been interpreted as showing that most of the accretion flows through
an accretion disc, but that part of the stream overflows the disc 
and so encounters the magnetosphere directly, giving rise to the 
beat-cycle modulation (e.g.\ Hellier 1991; 1993).  The idea that 
overflow occurs in cataclysmic variables is supported by theoretical 
modelling (e.g.\ Lubow 1989; Armitage \&\ Livio 1998). 

Murray et\,al.\ (1999) have argued against this idea as follows.  The
overflowing stream cannot penetrate further in than the radius of
minimum approach of a ballistic stream, $R_{\rm min}$, so cannot connect to
the magnetosphere unless $R_{\rm mag}$\,$>$\,$R_{\rm min}$; yet it is unclear whether
a disc can form if $R_{\rm mag}$\,$>$\,$R_{\rm min}$, since the stream could not orbit
freely around the magnetosphere. Murray et\,al.\ argue instead that
spiral shocks in the inner disc might give rise to orbital sidebands
of the spin frequency.

One counterargument is that a two-armed spiral would result in modulations 
at 2($\omega$\,--\,$\Omega$), whereas a stream leads to
$\omega$\,--\,$\Omega$, and we typically observe $\omega$\,--\,$\Omega$ 
rather than 2($\omega$\,--\,$\Omega$). A second
response is that in most IPs the magnetospheric radius is indeed
larger than $R_{\rm min}$.  As an example, take the case of AO~Psc, with
orbital and spin periods of 3.59 hr and 805 sec, respectively.
Deducing $R_{\rm mag}$ by assuming that the magnetosphere corotates with a
Keplerian disc at $R_{\rm mag}$, and choosing plausible stellar masses, we
find that $R_{\rm mag}$\,$\approx$\,1.2\,$\times$\,10$^{10}$\,cm, whereas
$R_{\rm min}$ is smaller at $\approx$\,0.6\,$\times$\,10$^{10}$\,cm.  Since
systems like AO~Psc do appear to accrete through discs, judging by the
dominance of the X-ray spin pulse, the task becomes explaining disc
formation in the regime where $R_{\rm mag}$\,$>$\,$R_{\rm min}$. Recent work on V2400~Oph
(Section~4) may shed light on this.

\section{FO~Aqr}
In turning now to individual systems, I first highlight FO~Aqr as 
the best-studied example of a system showing disc-overflow
accretion. The X-ray pulsation at the spin period is always largest,
but there is often (though not always) an additional pulsation at the
beat period (Hellier 1991; Norton et\,al.\ 1992a; Hellier 1993; 
Beardmore et\,al.\ 1998). The implication is that the overflow is
variable, occurring often but not always. 
FO~Aqr can be taken as an exemplar of the stars AO~Psc, V1223~Sgr \&\
BG~CMi, which have all shown a weak beat-cycle pulsation in at least
one X-ray observation (e.g.\ Hellier 1998).

\section{V2400~Oph}
V2400~Oph is at a low inclination of only $\approx$\,10\deg, so we
only ever see the `upper' magnetic pole (Buckley et\,al.\ 1995; Hellier
\&\ Beardmore 2002). As the stream flips to the hidden, lower pole,
the X-ray flux drops, resulting in a beat-cycle pulsation which dominates 
the X-ray lightcurves (Fig.~2). The idea is supported by optical spectroscopy,
where observed velocity changes over the beat cycle match well to a 
simulation of pole-flipping accretion, calculated for material 
5--10\,$R_{\rm wd}$ from the white dwarf (Fig.~3).    

\begin{figure}[t]   % Fig 2
\vspace*{4.4cm}
\includegraphics{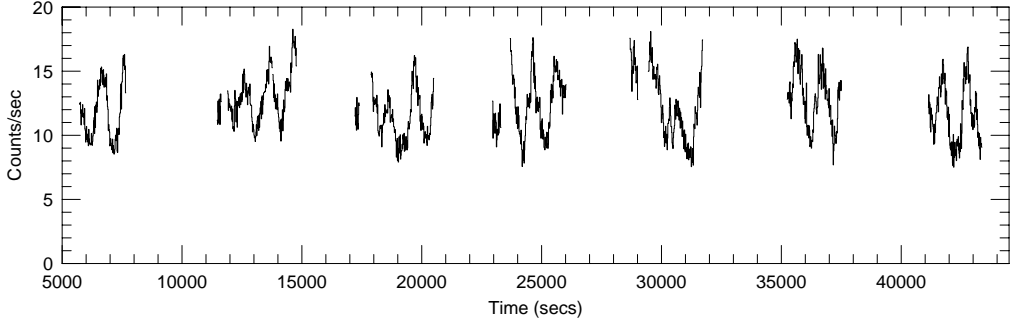}
\caption{Part of an {\sl RXTE\/} X-ray observation of V2400~Oph showing
an obvious beat-cycle pulsation. The Fourier transform of this observation
is shown in Fig.~1. (From Hellier \&\ Beardmore 2002.)}
\end{figure}

\begin{figure}[t]   % Fig 3
\vspace*{11.5cm}
\includegraphics{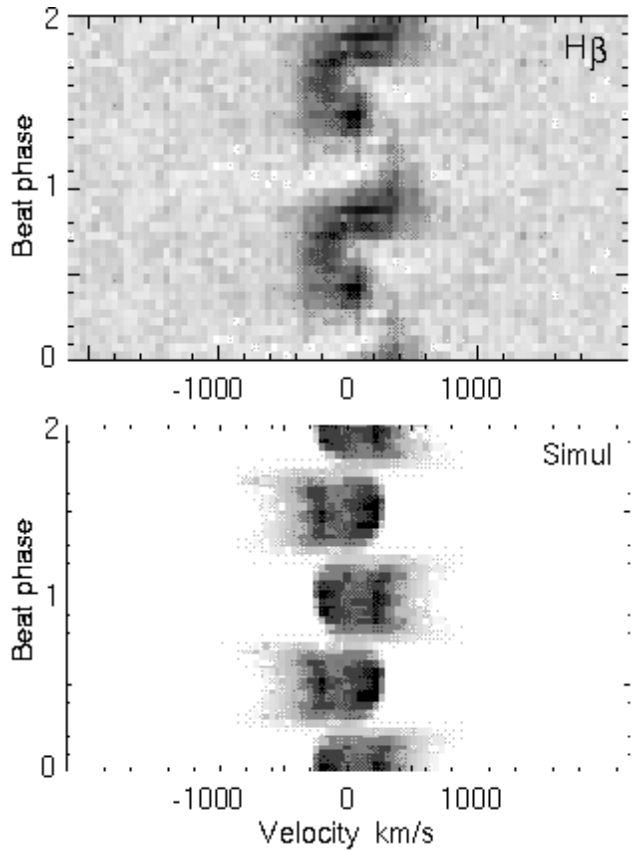}
\caption{The H$\beta$ line from V2400~Oph folded on the beat cycle,
and compared with a simulation of pole-flipping, stream-fed accretion
(see Hellier \&\ Beardmore 2002).} 
\end{figure}

However, a pole-flipping stream cannot be the whole story in V2400~Oph.
In Hellier \&\ Beardmore (2002) we concluded that much of the accretion flow 
is circling the white dwarf, not falling inward in a stream. 
 The evidence is, firstly, that the 
X-ray beat pulse is only 25\%\ deep. This implies that accretion is
always occuring at both poles continually, so that we still see accretion 
when the stream flips to the lower pole, suggesting that $\sim$\,75\%\  of
the flow does not participate in the pole flipping. Secondly, the 
velocity variation of the emission lines over the orbital cycle is only 
6 km\,s$^{-1}$. Even at 10\deg, the infall motion of a stream would mean that 
it would generate an orbital modulation of hundreds of km\,s$^{-1}$.
Only by diluting this with material circling the white dwarf is it 
possible to account for the observations. A third line of argument is that
while the changes in the line profile over spin and beat phases are
explained by stream-fed accretion (Figs.~3 \&\ 4), the 
varying components comprise  only $\approx$\,10\%\ of the total line line, 
implying that most line emission comes from something other than the stream. 

\begin{figure}[t]   % Fig 4
\vspace*{11.5cm}
\includegraphics{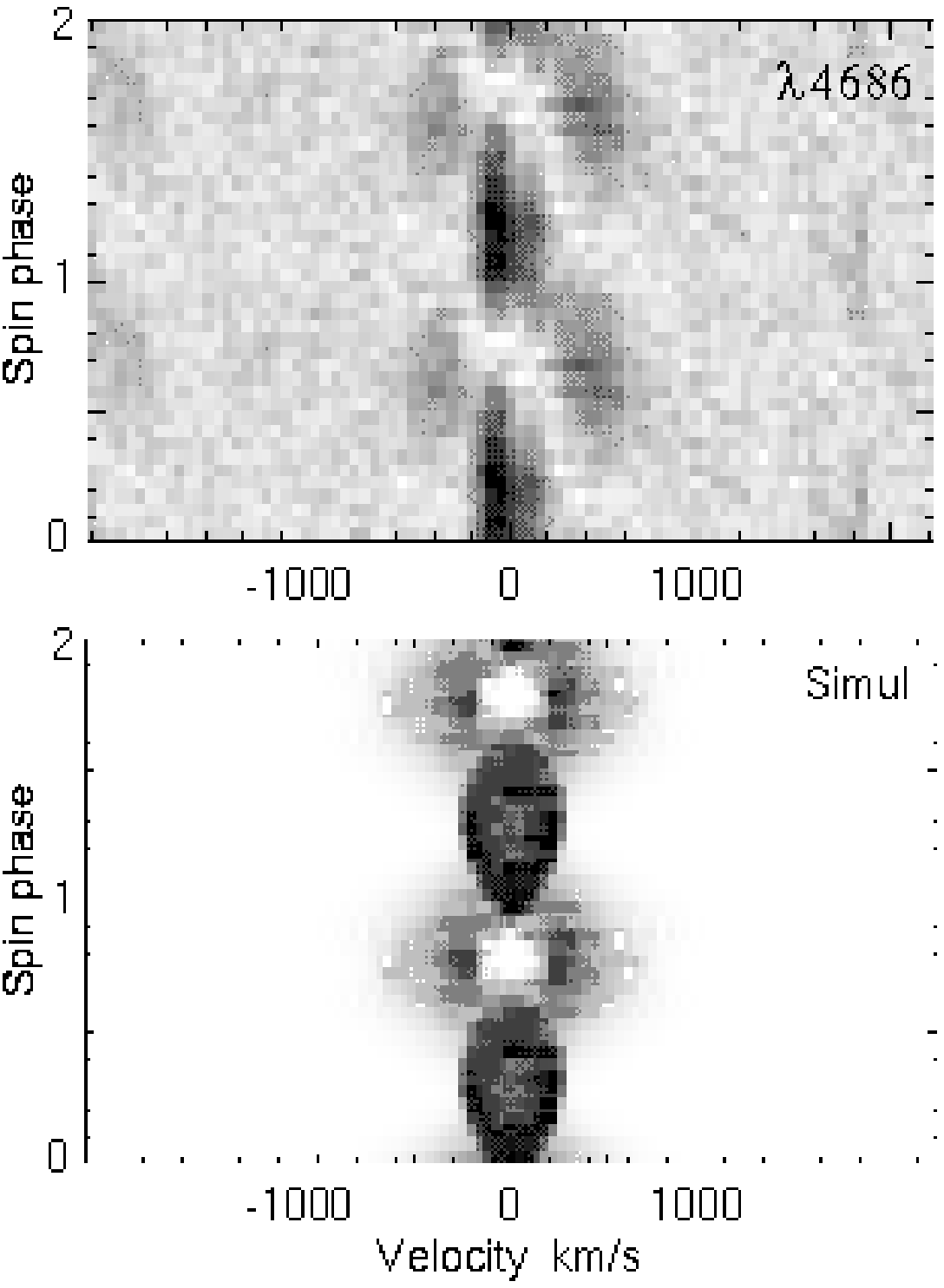}
\caption{The He\,{\sc ii}\,$\lambda$4686 line from V2400~Oph folded on 
the spin cycle, and compared with a simulation of stream-fed accretion 
(see Hellier \&\ Beardmore 2002).} 
\end{figure}

One possibility is that a disc is present, but this is unlikely
given the non-detection
of any X-ray spin pulse. Instead, we suggested (Hellier \&\ Beardmore 2002)
that the flow is in the form of diamagnetic blobs circling the white
dwarf, following the theoretical analysis of King (1993) and Wynn \&\ King 
(1995). The simulations by King \&\ Wynn (1999) show that, given a range of
blobs densities, the less-dense material is easily threaded and 
controlled by field lines, while denser blobs can cross field lines and orbit
the white dwarf.

The above picture may provide the answer to disc formation in the regime
$R_{\rm mag}$\,$>$\,$R_{\rm min}$, since it implies that dense blobs are crossing field
lines, losing memory of orbital phase, and spiralling inward to accrete
onto the white dwarf. If the field were somewhat lower, the drag on the
blobs would be reduced, slowing their inward spiral and allowing them to 
accumulate. They would tend to screen the field from one another, reducing
the effective field further, and so producing a runaway that would lead 
to disc formation. 

In this regard it is notable that V2400~Oph has the highest field
estimate among IPs at 9--27 MG (Buckley et\,al.\ 1995; V\"ath 1997), and
this might explain why it is discless. Even so, the presence of circling 
material suggests that it is on the edge of disc formation. A comparably 
high-field IP is PQ~Gem, with an estimate of 9--21 MG (V\"ath et\,al.\ 
1996; Potter et\,al.\ 1997), but this system appears
to be disc-fed, judging by the dominance of the X-ray spin pulse and
the lack on any X-ray beat pulse (Mason et\,al.\ 1992). Thus, the two
stars V2400~Oph and PQ~Gem could be delineating the conditions for
disc formation in an IP, with a slightly lower field and wider orbit
in PQ~Gem allowing a disc to form. 

\section{TX~Col}
Most IPs have X-ray lightcurves dominated either by the spin pulse or
by the beat pulse, allowing them to be interpreted as predominantly
disc-fed or predominantly stream-fed systems. In TX~Col, however, the two
pulsations are of comparable amplitude and are variable, so that the
spin-pulse is larger at some times and the beat pulse larger at others
(Buckley \&\ Tuohy 1989; Norton et\,al.\ 1997; Wheatley 1999).  One
possible explanation, as for FO~Aqr, is variable disc-overflow, but
this time with the flow through the stream being sufficient, on occasion,
to produce a beat pulse which is larger than the spin pulse.

However, it might also be possible to explain TX~Col along the lines
of V2400~Oph, with a combination of a stream and orbiting blobs. 
If a disc were on the verge of forming, the accretion flow could be hovering
between the two modes, explaining the variability of the X-ray
lightcurves. Arguing against this is the fact that polarisation is not
seen in TX~Col, suggesting that it is not among the higher-field
IPs. There has been relatively little optical work published on
TX~Col, and further spectroscopy may be the way to resolve the issue.

\section{V1025~Cen}
The two stars EX~Hya and V1025~Cen have exceptionally long spin periods 
with $P_{\rm spin}/P_{\rm orb}$\,$>$\,0.4, whereas in all other IPs the 
ratio is $<$\,0.1 (e.g.\ fig.~4 of Hellier, Beardmore, \&\ Buckley 1998). 
The condition $P_{\rm spin}/P_{\rm orb}$\,$<$\,0.1 is required for
equilibrium rotation with a disc (where the field's corotation velocity
equals the Keplerian velocity at $R_{\rm mag}$) since $R_{\rm mag}$ cannot
exceed the stream's circularisation radius, and this puts an upper
limit on $P_{\rm spin}$ (see, e.g., King \&\ Lasota 1991). One possible 
explanation is that EX~Hya and V1025~Cen do possess discs but are far 
from equilibrium, perhaps because of a drop in mass-transfer rate from 
which they are now recovering. 

A second possibility, though, has been suggested by King \&\ Wynn (1999).
They expound a discless model for these stars in which $R_{\rm mag}$ equals
the distance to the Lagrangian point.  Most of the mass-transfer stream 
accretes, but $\approx$\,10\%\ is expelled outward by the field to be
swept up by the secondary. The expulsion of high-angular-momentum
material balances the accretion torques and locks the system into an
equilibrium at much longer spin periods than in most IPs.

\begin{figure}[t]   % Fig 5
\vspace*{11cm}
\includegraphics{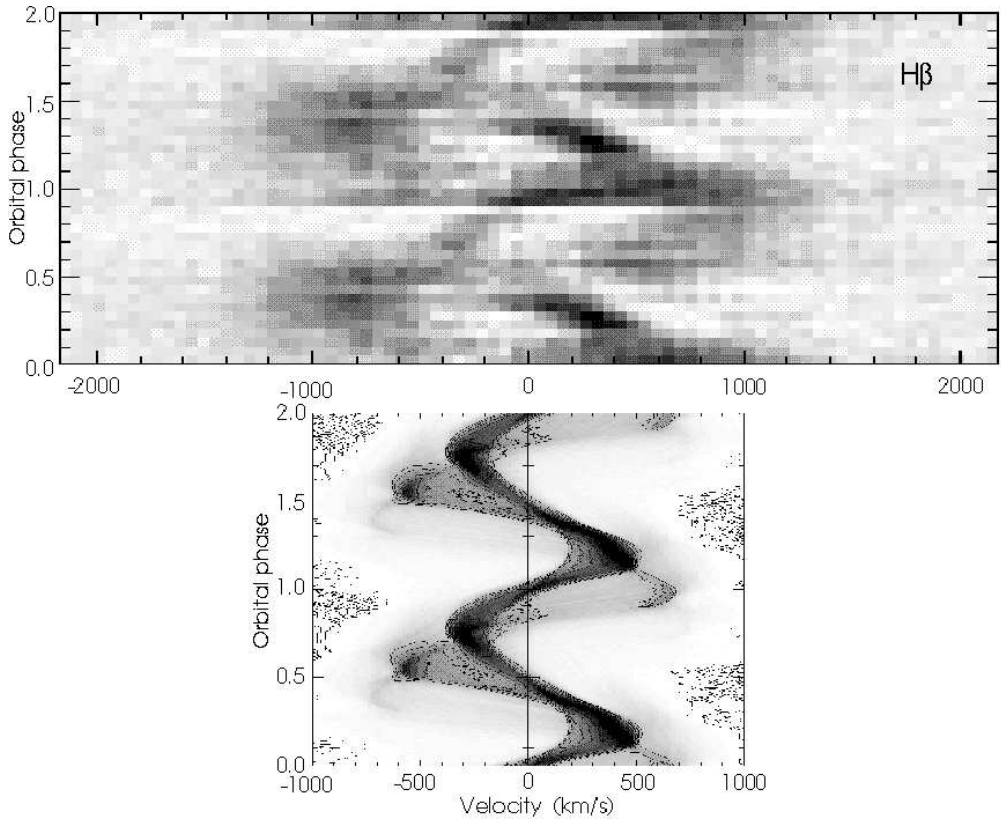}
\caption{The orbital variation of the H$\beta$ line of V1025~Cen 
({\it top\/}) on the same scale as a simulation by Graham Wynn 
({\it bottom\/}). The similarities include the central, lower-velocity 
S-wave, and a higher-velocity feature phased $\approx$\,0.25 cycles 
earlier. See Hellier et\,al.\ (2002) for a fuller account together 
with Doppler tomograms.}    
\end{figure}

In Hellier, Wynn, \&\ Buckley (2002) we present optical spectroscopy of
V1025~Cen aimed at deciding between the two models. We find that King \&\
Wynn's model has several successes in addition to accounting for the
anomalous spin period of this system.  In particular, there are
correspondences between the observed line profiles over the orbital 
cycle and the profiles predicted by the model (Fig.~5). Further, both
observed and model tomograms show the same `hook'-like features
(Wynn 2001; Hellier et\,al.\ 2002). 

However, the stream-fed model of King \&\ Wynn (1999) is again a
geometry varying with beat phase, and predicts a strong modulation
of accretion rate with beat phase (see their fig.~3). The X-ray
lightcurves, though, show only a spin-cycle pulsation (Hellier et\,al.\ 
1998). Although a spin-cycle pulse is readily explained even in a
stream-fed accretor, since views of the accretion sites vary with spin
phase, the absence of any beat-cycle pulsation is a problem for any
such model. A comparison of the Fourier transforms of X-ray
lightcurves of V2400~Oph and V1025~Cen would suggest that
the former is stream-fed and the latter disc-fed (Fig.~6).

\begin{figure}[t]   % Fig 6
\vspace*{6.5cm}
\includegraphics{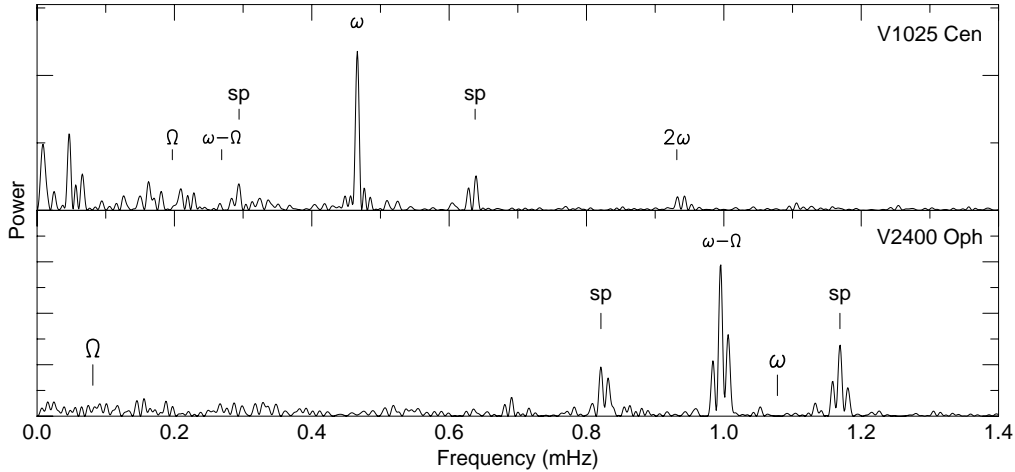}
\caption{A contrast of the X-ray Fourier transforms of V1025~Cen (which shows only
a spin-cycle pulse, suggesting that it is a disc-fed accretor) and
of V2400~Oph (which shows only a beat-cycle pulse, suggesting that
it is discless).}
\end{figure}

Thus, with points both for and against the two models, further work is
required to reach a conclusion about V1025~Cen. In particular, it is
worthwhile exploring whether a variant of the King \&\ Wynn model could
retain the same flow near the Lagrangian point, where expulsion of
material leads to the long spin period, but then give way to a disc-like
flow near the white dwarf, where dependence on orbital phase is washed out.

\section{EX~Hya}
Like V1025~Cen, the X-ray lightcurve of EX~Hya is always dominated by
a spin-cycle pulsation, pointing to disc-fed accretion.  A beat-cycle
pulsation has been seen only once, during an outburst (Hellier et\,al.\
2000). The eclipse profiles during outburst showed conclusively that the
accretion stream was greatly enhanced.  These findings fullfilled 
predictions made from an earlier outburst (Hellier et\,al.\ 1989) where
emission-line profiles suggested that the stream was overflowing the
disc and impacting on the magnetosphere. Thus, observations show that 
disc-overflow accretion occurs in EX~Hya during outburst (probably owing 
to enhanced mass-transfer) but not in quiescence.  It is currently
unclear whether the enhanced mass transfer is the sole cause of EX~Hya's
outbursts, or whether the enhancement is triggered by a disc instability
(see Hellier et\,al.\ 2000). If the latter, it would be proof of the 
presence of a disc. The quiescent studies show the double-peaked lines 
characteristic of an accretion disc, and an S-wave compatible with arising 
from the bright spot where the stream hits the disc (e.g.\ Hellier 
et\,al.\ 1987). Thus, overall the observations point to the presence of 
a disc in EX~Hya, despite the anomalous spin period. However, it should 
be noted that a detailed comparison of observations to the King \&\ Wynn 
(1999) model has not yet been published, and could develop this debate
further.

\section{AE Aqr}
To complete a round-up of stream-fed accretion, I mention AE~Aqr.  The
white dwarf in AE~Aqr is the fastest rotator among the secure IPs,
with a period of 33 sec. The current picture is that it has no disc, and 
the stream is expelled from the system by the propeller effect of the
rapidly spinning field (Wynn, King, \&\ Horne). The evidence is
primarily the finding that the spin-down energy of the white dwarf
exceeds the accretion energy (de Jager et\,al.\ 1994), suggesting that 
only a small fraction of the mass-transfer stream is accreted. The small
fraction that accretes is slowed by the propeller effect, presumably
circling the white dwarf many times before accreting; it loses
dependence on orbital phase and thus produces pulses at the spin
period, although given the slow infall they are seen in soft
X-rays only. Again, a detailed comparison of the observations to this
model has yet to be published, so it is too early for final
conclusions, but see Welsh (1999) for a review of the state of play.

\end{document}